\documentclass[a4paper,11pt]{article}
\usepackage{jcappub}
\usepackage{booktabs}

\newcommand{\Xfirst}{X_\mathrm{first}}
\newcommand{\Xmax}{X_\mathrm{max}}
\newcommand{\Xzero}{X_0}
\newcommand{\DX}{\Delta X}
\newcommand{\Erec}{E_\mathrm{rec}}
\newcommand{\dEdX}{\mathrm{d}E_\mathrm{dep}/\mathrm{d}X}
\newcommand{\lgE}{\log_{10}(E/\mathrm{eV})}

\newcommand{\gcm}{\ensuremath{\mathrm{g\,cm^{-2}}}}

\newcommand{\Conex}{\textsc{Conex}}
\newcommand{\Sibyll}{\textsc{Sibyll}~2.3d}
\newcommand{\Epos}{EPOS~LHC-R}

\title{Reconstructing the depth of first interaction in extensive air showers from fluorescence detector profiles with a physics-motivated neural network}

\author[1]{Marcos Olegario\note{Corresponding author.}}
\author{and Vitor de Souza}

\emailAdd{marcos.olegario@usp.br}
\emailAdd{vitor@ifsc.usp.br}

\affiliation{Instituto de F\'isica de S\~ao Carlos, Universidade de S\~ao Paulo, Av. Trabalhador S\~ao-carlense 400, S\~ao Carlos, Brasil}

\abstract{We present a physics-motivated neural network that reconstructs, event by event, the depth of the first interaction of ultra-high-energy cosmic rays from longitudinal profiles measured by fluorescence detectors. The network couples a convolutional encoder to a decoder that implements the Gaisser--Hillas function and predicts quantiles of the target, providing a per-event uncertainty estimate. It is trained and evaluated on four million proton- and iron-induced showers between $10^{18}$ and $10^{20}$~eV, simulated with the hadronic interaction models EPOS LHC-R and Sibyll~2.3d and processed by an end-to-end simulation of the fluorescence detector response. Despite detector effects, the reconstruction retains useful information for protons, achieving a mean absolute error of about 21~\gcm\ for primary energies around $10^{19}$~eV,  whereas for iron most of the information is lost.  The reconstructed distributions preserve the exponential shape of the true ones, so the proton--air interaction length can be recovered from their tail with a model-related systematic uncertainty of approximately 4\%. This value is limited mainly by the choice of the hadronic interaction model and is nearly insensitive to iron contamination. These results show that the depth of the first interaction can be reconstructed under realistic detection conditions, offering a complementary route to measurements of the proton--air cross section and mass-composition studies.}

\keywords{ultra high energy cosmic rays, cosmic ray experiments, machine learning}

\hypersetup{pdftitle={Reconstructing the depth of first interaction in extensive air showers from fluorescence detector profiles with a physics-motivated neural network},pdfauthor={Marcos Olegario, Vitor de Souza}}

\begin{document}
\maketitle
\flushbottom

\section{Introduction}
\label{sec:intro}

Among the properties of ultra-high-energy cosmic rays (UHECRs), charged particles with energies above $10^{18}$ eV, the characterization of their mass composition remains an open problem in astroparticle physics~\cite{kampert2012}. Fluorescence detectors measure the longitudinal development of extensive air showers (EAS) initiated by UHECRs in the atmosphere, from which the depth of shower maximum, $\Xmax$, can be determined~\cite{souza2017}. The distribution of $\Xmax$ is used to investigate the mass composition of UHECRs~\cite{bellido_2017,batista_2019} and to constrain the proton--air cross section~\cite{PAcross2012,TAcross2015}. However, $\Xmax$ is intrinsically affected by fluctuations arising from both the depth of the first interaction of the primary particle, $\Xfirst$, and the subsequent shower development, characterized by $\DX = \Xmax - \Xfirst$. The latter has been modeled using simulations~\cite{ulrichMeasurementProtonairCross2009}, but its fluctuations, driven by the stochastic nature of shower development, weaken the sensitivity of $\Xmax$ to the properties of the primary particle.

The depth of the first interaction is more directly related to the primary particle. For a primary of mass $A$ and energy $E_0$, $\Xfirst$ follows an exponential distribution with mean $\lambda_{A\text{-air}}(E_0) = \langle m_\mathrm{air}\rangle/\sigma_{A\text{-air}}(E_0)$, where $\langle m_\mathrm{air}\rangle$ is the mean target mass of air~\cite{ulrichMeasurementProtonairCross2009}. This distribution provides access to nucleus--air cross sections at center-of-mass energies well beyond those achievable at particle accelerators~\cite{TAcross2015}. Reconstructing $\Xfirst$ can therefore provide an estimate of this distribution, offering a complementary approach to cross-section measurements and mass-composition studies.

Estimating $\Xfirst$ for individual showers is, however, far from straightforward. The first interaction takes place high in the atmosphere, where the early shower development produces insufficient fluorescence light for detection, requiring $\Xfirst$ to be inferred from the subsequent profile. Ref.~\cite{andringaMassCompositionCrosssection2011} showed that the shape of the longitudinal profile is correlated with $\DX$, enabling event-by-event estimates of $\Xfirst$. Previous studies have applied machine-learning techniques to estimate $\Xfirst$ from features extracted from longitudinal profiles~\cite{yapici_2015}. 
However, these studies relied on profiles obtained directly from shower simulations. The detector response was either neglected, applied a posteriori, or approximated through fixed truncations of the observed depth range. Consequently, the reconstruction of $\Xfirst$ under realistic observation conditions remains to be investigated.

Beyond the reconstruction of $\Xfirst$, machine learning methods are increasingly used in cosmic-ray physics \cite{abdulhalim_2025,Auger2021XmaxDNN,wang_2026}, yet conventional approaches often provide limited physical interpretability and point estimates without quantifying predictive uncertainty. Moreover, their learned representations are not necessarily associated with interpretable physical quantities. Physics-informed machine learning addresses these limitations by incorporating physical knowledge into network architectures or training procedures~\cite{kashinath2021}. For example, autoencoder latent variables can be constrained to represent physical-model parameters~\cite{yang2022}, linking data-driven predictions to interpretable quantities. To our knowledge, this approach has not yet been applied to air-shower profile reconstruction.

In this work, we present a physics-motivated neural network that reconstructs $\Xfirst$ event by event from the longitudinal profiles. To obtain realistic inputs, the detector response was simulated, accounting for the emission, propagation, and detection of fluorescence light, the event reconstruction, and quality cuts similar to those applied to real data. The network is an autoencoder whose decoder implements the Gaisser--Hillas function, so that part of its latent representation corresponds to physical parameters of the profile. In addition to estimating $\Xfirst$, the network provides event-level uncertainty estimates while preserving the exponential shape of the $\Xfirst$ distribution. The method is applied to proton- and iron-induced showers between $10^{18}$ and $10^{20}$~eV simulated with two hadronic interaction models, without informing the network of the primary type. We quantify the information lost in the detection process by comparing detector-level with ideal profiles, and the dependence of the reconstruction on the hadronic interaction model by training and testing the network on different models.

This paper is organized as follows. Section~\ref{sec:data} describes the generation of simulated extensive air-shower datasets. Section~\ref{sec:model} presents the deep-learning methodology, focusing on the network architecture, loss function, and training procedure. Section~\ref{sec:res1} presents and discusses the model's performance in event-by-event reconstruction, while Section~\ref{sec:res2} investigates its application to estimating the proton--air mean free path. Finally, Section~\ref{sec:conc} summarizes the main conclusions and outlines future directions.

\section{Dataset simulation and pre-processing}
\label{sec:data}

The dataset used to train and evaluate the model comprises $4 \times 10^6$ air showers simulated with \Conex{}~\cite{conex1,conex2}. Four libraries of $10^6$ events each were produced, covering proton and iron primaries and the hadronic interaction models \Epos{}~\cite{pierogEPOSLHCRGlobalApproach2025} and \Sibyll{}~\cite{SIBYLL2020}. Primary energies were sampled uniformly in $\lgE$ between $10^{18}$ and $10^{20}$~eV, and all showers were simulated at a zenith angle of $60^\circ$. This inclination provides a sufficiently large slant depth to simulate nearly the full longitudinal development of each shower.

The response of a fluorescence detector to these showers was simulated with the framework described in Ref.~\cite{arapy_zenodo}. For each shower, a geometry was drawn, with a core position sampled uniformly within a radius of 50~km around the detector, a new zenith angle sampled in $[0^\circ, 60^\circ]$ uniformly in $\sin^2\theta$, and an azimuth angle drawn uniformly in $[0^\circ, 360^\circ)$.  The $60^\circ$ zenith angle used in the \Conex{} simulations was chosen to ensure sufficiently long longitudinal profiles and does not constrain the zenith angles of the detection geometries. The \Conex{} profiles, expressed as energy deposit per unit slant depth, were then mapped onto the shower axes defined by these new geometries. Specifically, the slant depths were converted into positions along each axis using the atmospheric model of the framework. The fluorescence photons produced along the resulting shower axes were then propagated to the detector, accounting for atmospheric scattering and attenuation. The simulation also includes the electronic noise of the detector, the trigger logic, and the event reconstruction, which yields the reconstructed longitudinal profile and primary energy. Only the portions of the profiles above ground and within the field of view contribute to the detector signal.

The detector configuration was modeled after the fluorescence telescopes of the Pierre Auger Observatory~\cite{abreuExposureHybridDetector2011, abrahamFDAuger2010}: an aperture area of 3.8~m$^2$, a field of view spanning elevations from $1.5^\circ$ to $30^\circ$, a camera with $1.5^\circ$ pixels, and an altitude of 1400~m above sea level. A total (optical and quantum) efficiency of 20\% was assumed~\cite{carvalhoEffectsEnergyError2007}. Atmospheric conditions and fluorescence emission were modeled with the default settings of Ref.~\cite{arapy_zenodo}, including its atmospheric density profile, aerosol model, and fluorescence yield.

The detected profiles were subjected to quality cuts similar to those of Ref.~\cite{thepierreaugercollaborationMeasurementAverageShape2018}. The observed track had to span at least 300~\gcm\ and contain the reconstructed $\Xmax$. To reproduce the geometric selection applied to real data, where it suppresses direct Cherenkov contamination, the angle between the shower axis and the viewing direction of every pixel with signal had to exceed $20^\circ$. In addition, at least 200~\gcm\ had to be observed before $\Xmax$, since the rising edge of the profile carries most of the information on $\Xfirst$~\cite{andringaMassCompositionCrosssection2011}. 

\begin{figure}[htbp]
    \centering
    \includegraphics[width=1\linewidth]{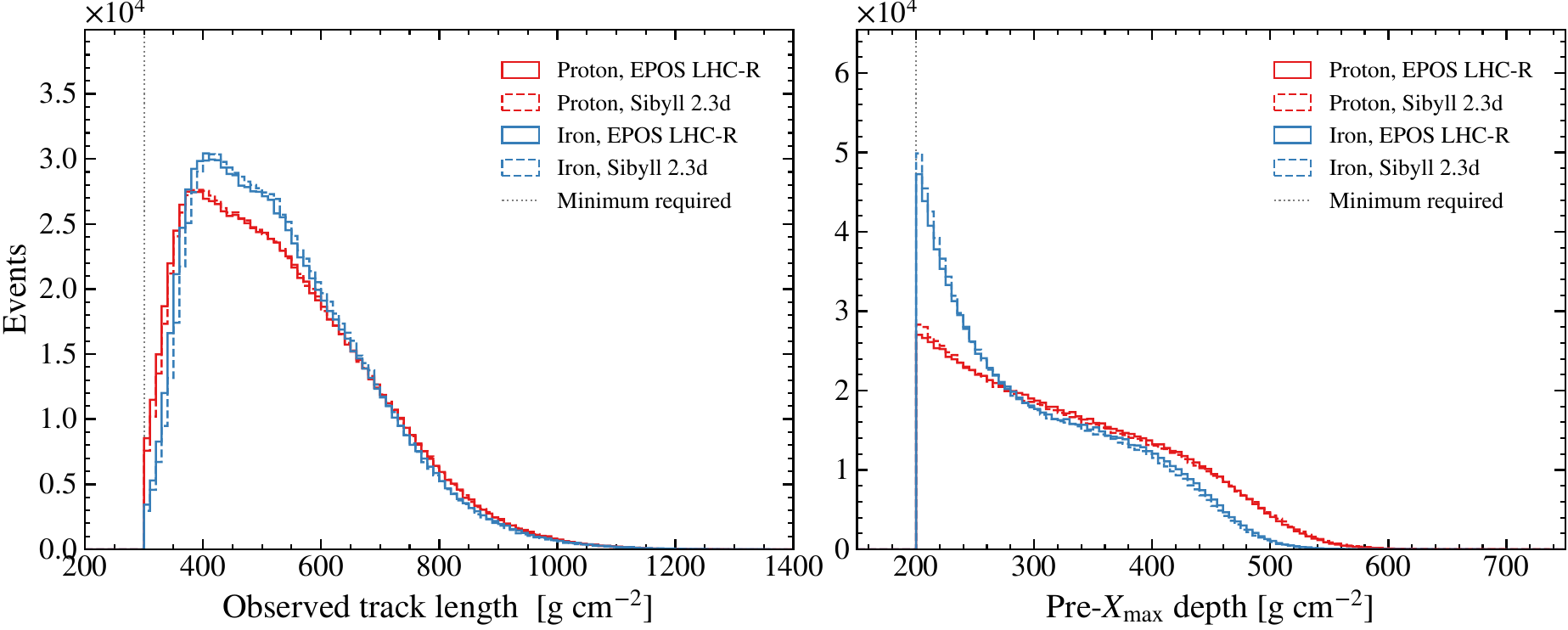}
    \caption{Distributions of the observed track length (left) and of the depth observed before $\Xmax$ (right) for the events in each library of detector-level profiles, after all quality cuts. The dotted vertical lines indicate the minimum values required by the selection.}
    \label{fig:cuts}
\end{figure}

For each shower, geometries were redrawn until the event triggered and passed all cuts. This procedure is not intended to reproduce the population of events selected in measured data. Instead, it ensures that all simulated showers contribute to the training and evaluation datasets, providing a controlled test of the reconstruction in which the original $\Xfirst$ distribution is preserved. The dependence of the selection efficiency on $\Xfirst$ and its implications for the proton--air interaction length measurement are discussed in Section~\ref{sec:res2}. Figure~\ref{fig:cuts} shows the distributions of the observed track length and of the depth observed before $\Xmax$.

\begin{figure}[htbp]
    \centering
    \includegraphics[width=0.75\linewidth]{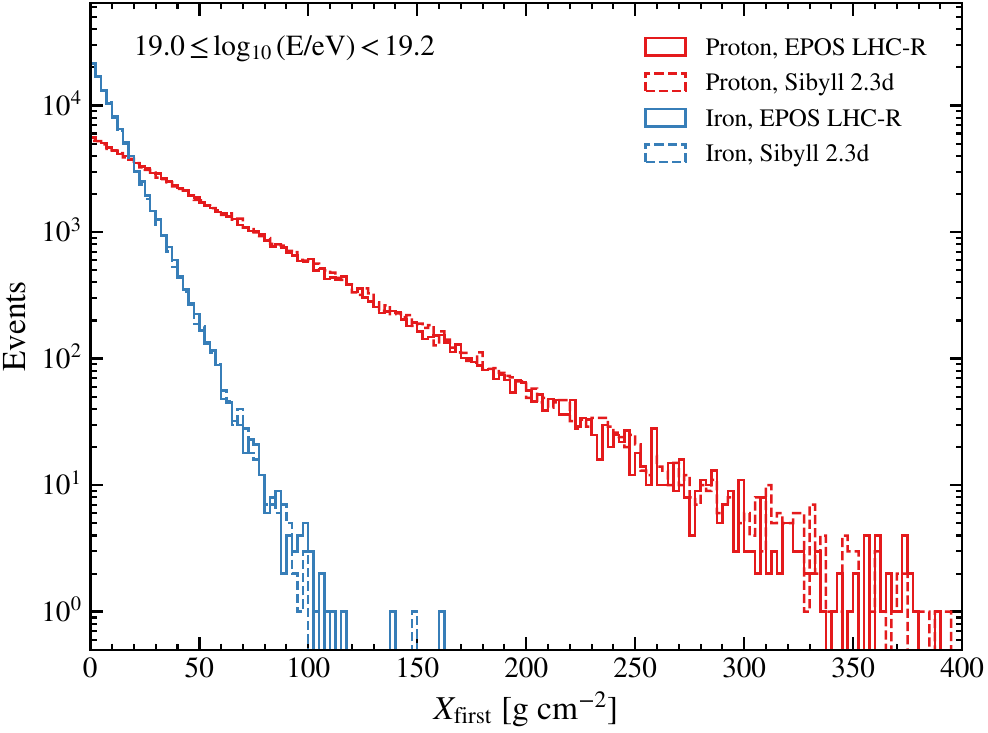}
    \caption{Distributions of the depth of the first interaction for proton and iron primaries simulated with \Epos{} and \Sibyll{}, shown for $19.0 \leq \lgE < 19.2$.}
    \label{fig:xfirst_dist}
\end{figure}

After the detector simulation, the following quantities were stored for each event: the observed longitudinal profile, i.e.\ the slant depths $X_i$ and the corresponding energy deposits $(\dEdX)_i$, and the reconstructed primary energy $\Erec$. We also stored the parameters of the Gaisser--Hillas (GH) function~\cite{gaisserhillasref},
\begin{equation}
  \frac{\mathrm{d}E_\mathrm{dep}}{\mathrm{d}X}(X) =
  \left(\frac{\mathrm{d}E_\mathrm{dep}}{\mathrm{d}X}\right)_{\!\mathrm{max}}
  \left(\frac{X-\Xzero}{\Xmax-\Xzero}\right)^{\frac{\Xmax-\Xzero}{\lambda}}
  \exp\!\left(\frac{\Xmax-X}{\lambda}\right),
  \label{eq:gh}
\end{equation}
fitted to the measured points, namely $(\dEdX)_\mathrm{max}$, $\Xmax$, and $\Xzero$, with $\lambda$ fixed at 70~\gcm. We note that $\Xzero$ is a shape parameter of the fit with no direct physical relation to the depth of the first interaction, and it may even take negative values. Finally, we stored the depth of the first interaction, $\Xfirst$, which is the target of the model. Figure~\ref{fig:xfirst_dist} shows the $\Xfirst$ distributions for each primary and hadronic interaction model in the energy range of $10^{19.0-19.2}$~eV.

The detected events were grouped into three datasets according to the hadronic interaction model: one containing only \Epos{} events, one containing only \Sibyll{} events, and one combining both, hereafter referred to as \textit{Mix}.
Each dataset contains equal numbers of proton- and iron-induced showers, so the network is not informed of the primary type.

To quantify the information lost in the detection process, we built a second set of libraries from the same showers and with the same partition, using the profiles produced directly by \Conex{} instead of the detector-level ones. These libraries, hereafter referred to as \emph{ideal}, contain the slant depths and energy-deposit profile in steps of 10~\gcm, the true primary energy, the GH parameters, and $\Xfirst$. Since a \Conex{} profile starts at the first interaction, its onset alone would reveal $\Xfirst$. To prevent this leakage, each profile is restricted to the range $(X - \Xmax^{\rm sim}) \in [-400,100]$~\gcm~\cite{andringaMassCompositionCrosssection2011}, where $\Xmax^{\rm sim}$ is the true shower maximum extracted from \Conex{}.

Before training, the data were preprocessed as follows. The events of each dataset were split into training (70\%), validation (15\%), and test (15\%) subsets, stratified by primary and by bins of simulated energy.
Unlike the ideal profiles, detector-level profiles have variable length and nonuniform sampling.
They were therefore zero-padded to 100 bins and accompanied by a binary mask that flags the measured points.
The energy deposit was expressed as $\log_{10}(\dEdX)$ to compress its dynamic range, which spans several orders of magnitude.
The target was transformed to $\log_{10}(1+\Xfirst)$, with $\Xfirst$ in \gcm, which reduces the skewness of its exponential distribution and avoids the logarithm of zero. 

\begin{figure}[htbp]
    \centering
    \includegraphics[width=1\linewidth]{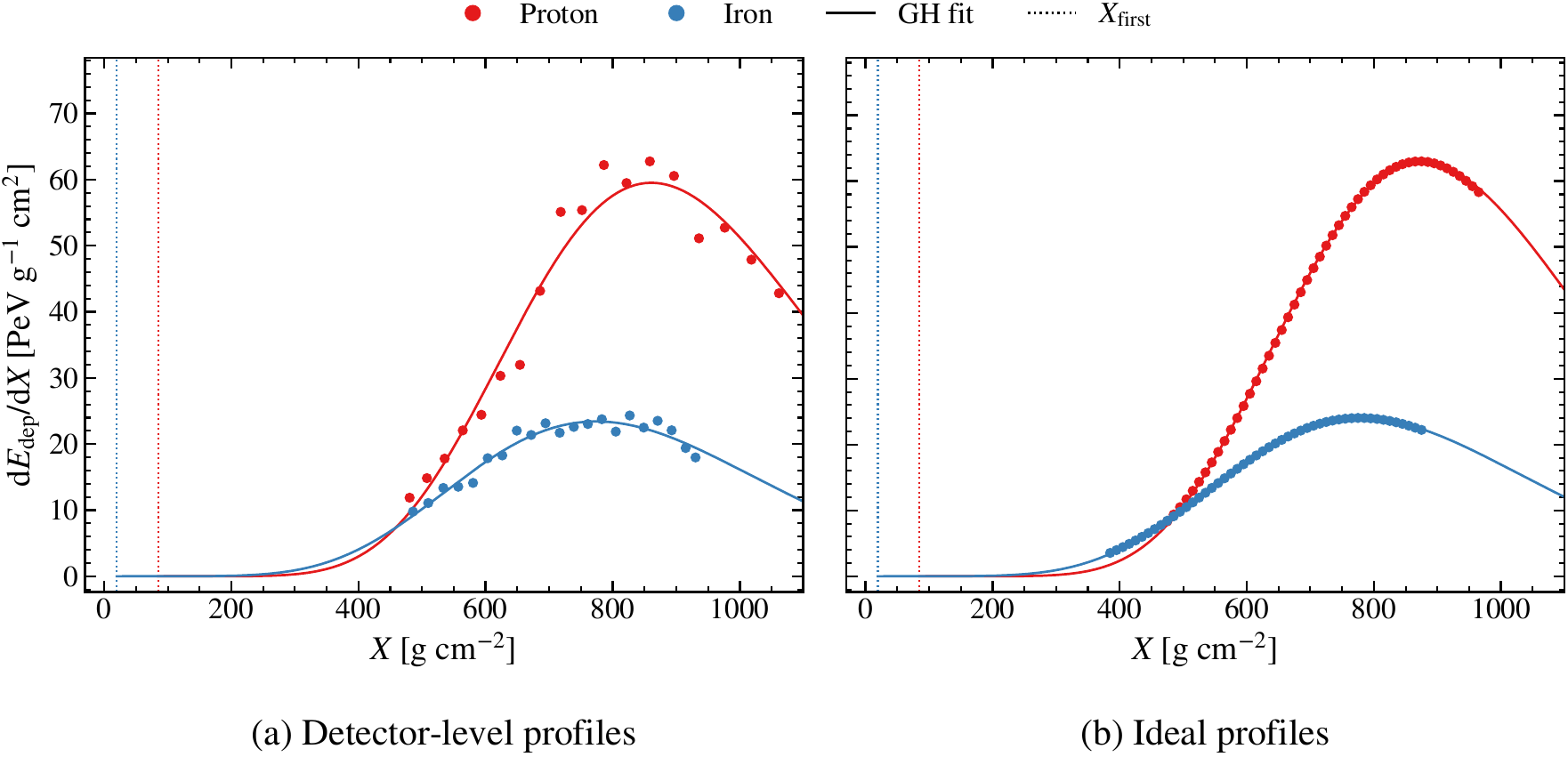}
    \caption{Examples of detector-level and ideal longitudinal profiles for proton- and iron-induced showers generated with \Epos{}, with primary energies of $10^{19.5}$ and $10^{19}\,\mathrm{eV}$, respectively. A Gaisser--Hillas fit to each profile is shown. Profiles are displayed in linear scale for clarity, although the network operates on $\log_{10}(\dEdX)$.}
    \label{fig:profiles}
\end{figure}

Because the input features differ by many orders of magnitude, they were rescaled to contribute comparably to the gradient updates: profile values, slant depths, GH parameters, and energies were min--max normalized to $[0,1]$, and the transformed target was standardized to zero mean and unit variance. The normalization parameters were computed from the training subsets, separately for ideal and detector-level libraries, and applied unchanged to validation and test subsets. No information from the test data therefore enters the preprocessing. Figure~\ref{fig:profiles} shows examples of profiles from both libraries.

\section{Physics-motivated neural network}
\label{sec:model}

A central challenge in this work was to design a neural network that explicitly incorporates physical knowledge into the reconstruction of $\Xfirst$. Two features of the problem guided this design. First, the longitudinal shower profile has a well-known shape, accurately described by the Gaisser--Hillas function of Eq.~\eqref{eq:gh}. Second, $\Xfirst$ follows an exponential distribution (Fig.~\ref{fig:xfirst_dist}) and is only partially constrained by the profile. The architecture therefore embeds the GH function in the reconstruction pipeline, so that the network must describe each profile through physically meaningful parameters. In addition, it estimates the conditional distribution of $\Xfirst$ instead of a single value per event. The spread of the predictions then reflects the fluctuations of the shower development, and combining the predicted distributions of all events recovers the exponential shape of the $\Xfirst$ distribution, whose slope is related to the proton--air cross section. This section describes how these two elements were incorporated into the network.

\subsection{Architecture}
\label{sec:architecture}

\begin{figure}[htbp]
  \centering
  \includegraphics[width=\textwidth]{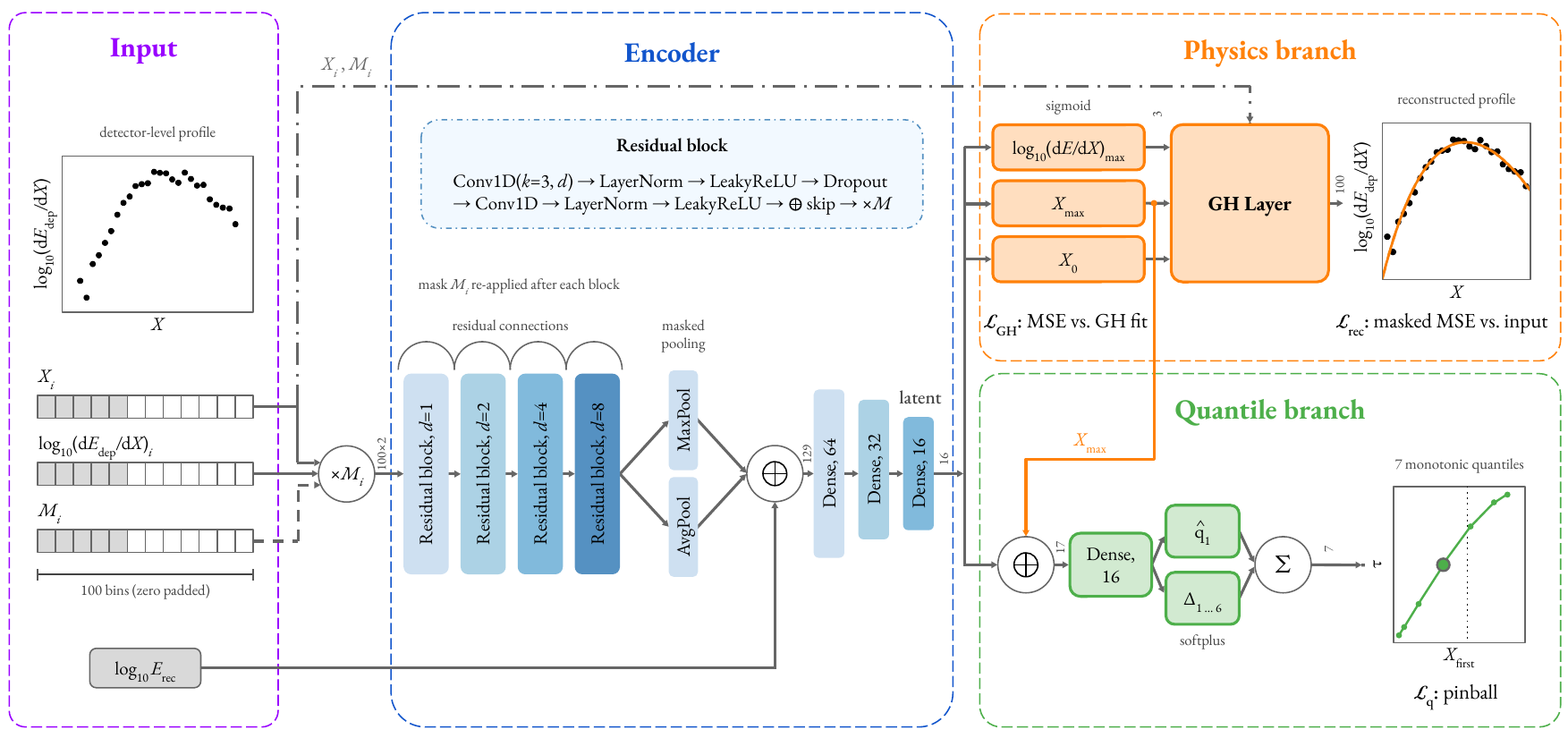}
  \caption{Network architecture for reconstructing $X_\mathrm{first}$. Insets show a proton shower simulated with \Epos{} at $\log_{10}(E/\mathrm{eV}) = 19.3$ from the test subset, reconstructed by the Mix network (Section \ref{sec:training}). They show the input profile, its GH reconstruction, and the predicted quantiles, with the predicted median highlighted and the true $X_\mathrm{first}$ indicated by the dashed line. Gray numbers denote tensor dimensions, and loss terms are indicated beside their corresponding outputs.}
  \label{fig:architecture}
\end{figure}

The architecture, shown in Figure~\ref{fig:architecture}, was inspired by the Autoencoder with Latent Physics (ALPS)~\cite{yang2022}. In ALPS, an estimator infers the free parameters of a known physical model, and the model itself generates the reconstruction that is compared with the observations. We adapted this idea to our problem: a convolutional encoder maps the profile to a latent representation, which feeds two output branches. In the physics branch, the GH function takes the role of the physical model, while the quantile branch, which has no counterpart in ALPS, estimates the distribution of $\Xfirst$.

The inputs are the slant depths $X_i$ and energy deposits $\log_{10}(\mathrm{d}E_\mathrm{dep}/\mathrm{d}X)_i$ of the profile, the mask that flags the measured points, and the reconstructed energy $\log_{10}(E_\mathrm{rec}/\mathrm{eV})$, all preprocessed as described in Section~\ref{sec:data}. The two profile sequences are concatenated and processed by a temporal convolutional network (TCN)~\cite{bai2018}, composed of four residual blocks~\cite{he2015} with 64 filters, kernel size 3, and dilation rates of 1, 2, 4, and 8. Each block applies two dilated convolutions, each followed by layer normalization~\cite{ba2016} and a LeakyReLU activation, with a dropout rate of 0.05 between them. The mask is applied to the input and to the output of every block, so that the padded bins never contribute to the result. 
Masked global average and maximum pooling then reduce the sequence to 128 features, independently of the number of measured points, and the energy is appended to this vector. Three fully connected layers of 64, 32, and 16 units with LeakyReLU activations map the result to a 16-dimensional latent representation, which is shared by the two output branches.

The physics branch predicts the GH parameters $\log_{10}(\mathrm{d}E_\mathrm{dep}/\mathrm{d}X)_\mathrm{max}$, $X_\mathrm{max}$, and $X_0$ through three units with sigmoid activation, whose outputs lie in the normalized range of each parameter. A decoder layer with no trainable weights maps these values back to physical units and evaluates Eq.~\eqref{eq:gh}, with $\lambda = 70~\gcm$, at the measured depths $X_i$ of the event. The result is a reconstructed profile in the same representation as the input. This branch forces the latent representation to encode a profile consistent with the GH parametrization and makes the bottleneck of the branch physically interpretable, since each of its outputs is a GH parameter. In addition, as discussed in Section~\ref{sec:loss}, the comparison between the reconstructed and input profiles provides a training signal that does not rely on simulated quantities.

Unlike the GH parameters, however, $\Xfirst$ is not fully determined by the profile. The stochastic nature of air-shower development limits the reconstruction of $\Xfirst$ from the longitudinal profile alone. Fluctuations in the first hadronic interactions make the distance between $\Xfirst$ and $X_\mathrm{max}$ vary considerably from shower to shower~\cite{ulrichMeasurementProtonairCross2009}, and the shape of the profile only partially constrains it~\cite{andringaMassCompositionCrosssection2011}. As a consequence, a given profile is compatible with a range of values of $\Xfirst$. A single estimate per event discards this spread and tends to concentrate the reconstructed values around the mean, distorting the exponential shape of the reconstructed distribution~\cite{yapici_2015}. Our network therefore estimates, for each event, the conditional distribution of $\Xfirst$ given the observed profile, through seven quantiles predicted by the quantile branch.

The quantile branch receives the latent vector concatenated with the predicted $X_\mathrm{max}$. This direct connection exploits the correlation between the two depths. A hidden layer of 16 units feeds two outputs: the lowest quantile $\hat q_1$ and six increments $\Delta_l = \mathrm{softplus}(\cdot) > 0$. The quantiles are then obtained as
\begin{equation}
  \hat q_k = \hat q_1 + \sum_{l=1}^{k-1} \Delta_l, \qquad k = 1, \dots, K,
  \label{eq:monotonic}
\end{equation}
which makes them monotonic by construction and prevents the quantile crossing that affects independently estimated quantiles~\cite{chernozhukov2010,cannon2018}. We use $K = 7$ levels, $\tau_k \in \{$0.05, 0.10, 0.25, 0.50, 0.75, 0.90, 0.95$\}$, defined in the space of the standardized target (Section~\ref{sec:data}). Since this transformation is monotonic, the predicted quantiles are converted back to $\gcm$ without changing their levels. The median $\hat{q}_{0.50}$ is taken as the point estimate of $\Xfirst$.

The network was implemented in TensorFlow \cite{tensorflow2015} using the Keras API \cite{chollet2015keras} and comprises 99\,514 trainable parameters, 89\% of which belong to the TCN. Hyperparameter selection was performed using a combination of random search and grid search \cite{bergstra2012}, exploring a wide range of configurations and selecting those that yielded the lowest validation loss. The same architecture was used for the ideal profiles, with the true primary energy provided as input.

\subsection{Loss function}
\label{sec:loss}

The network was trained by minimizing a weighted sum of one loss per output,
\begin{equation}
  \mathcal{L} = w_\mathrm{GH}\,\mathcal{L}_\mathrm{GH} + w_q\,\mathcal{L}_q + w_\mathrm{rec}\,\mathcal{L}_\mathrm{rec}.
  \label{eq:loss}
\end{equation}
The weights were not fixed during training, and their schedule is described in Section~\ref{sec:training}.

\begin{figure}[htbp]
  \centering
  \includegraphics[width=0.85\textwidth]{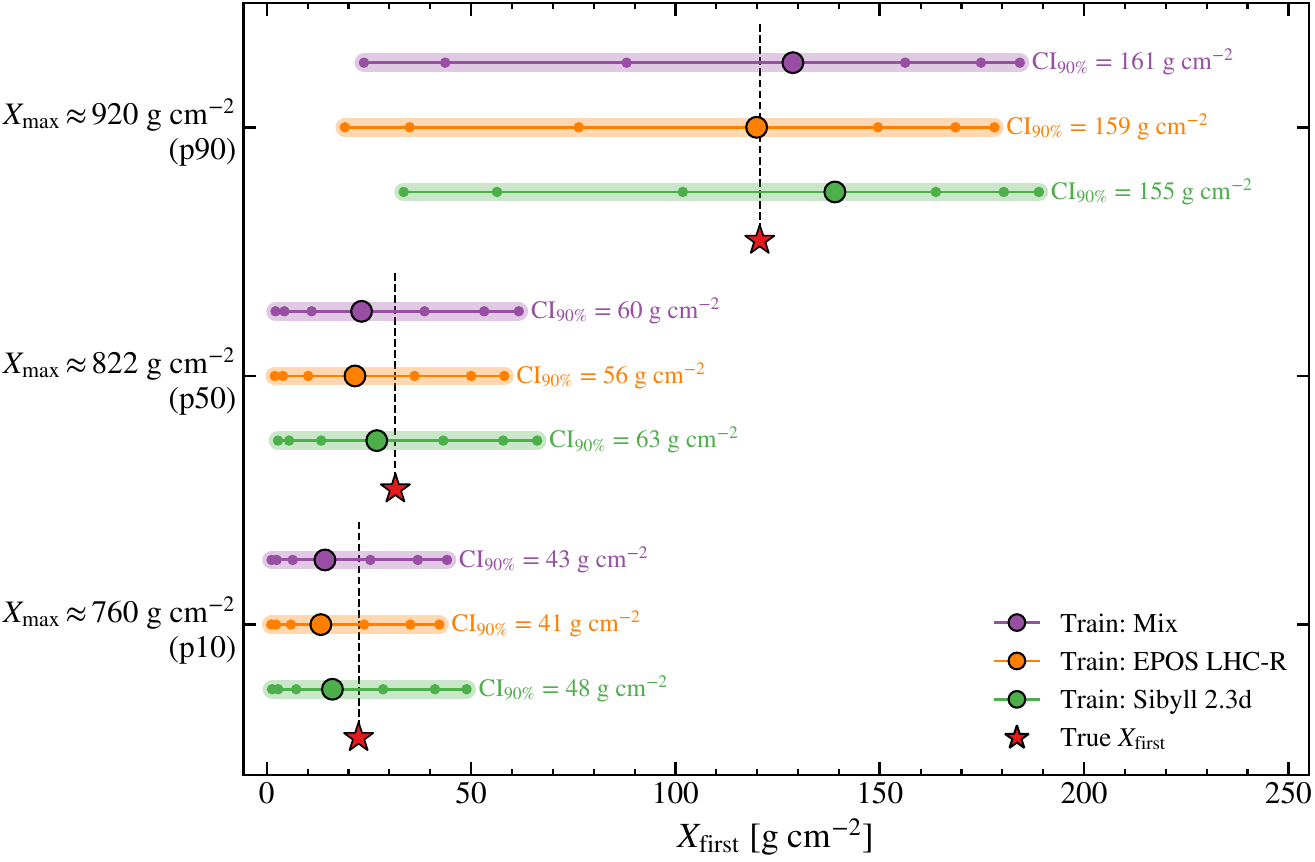}
  \caption{Predicted quantiles of $\Xfirst$ for three proton-induced longitudinal profiles from the \Epos{} test dataset, using networks trained on the three detector-level datasets as described in Section~\ref{sec:training}. The profiles correspond to the 10th (p10), 50th (p50), and 90th (p90) percentiles of the dataset's $\Xmax$ distribution. The star marks the true $X_\mathrm{first}$, the large circle the predicted median, and the small circles the predicted quantiles. The outermost quantiles delimit the central $90\%$ interval.}
  \label{fig:examplesCI}
\end{figure}

The first term, $\mathcal{L}_\mathrm{GH}$, is the mean squared error between the predicted GH parameters and those fitted to the profile (Section~\ref{sec:data}). The second is the pinball loss~\cite{koenker1978,koenker2005}, averaged over the $K$ quantile levels,
\begin{equation}
  \mathcal{L}_q = \frac{1}{K}\sum_{k=1}^{K} \max\!\big[\tau_k\,(y-\hat q_k),\ (\tau_k-1)\,(y-\hat q_k)\big],
  \label{eq:pinball}
\end{equation}
where $y$ is the standardized target and $\hat q_k$ is the quantile predicted at level $\tau_k$. The expected pinball loss is minimized by the conditional quantile of level $\tau_k$, so the seven outputs approximate the conditional distribution of $\Xfirst$ given the profile and the energy. The median $\hat q_{0.50}$ is used as the point estimate, $\Xfirst^{\rm pred}$, and $[\hat q_{0.05}, \hat q_{0.95}]$ defines a per-event prediction interval with a nominal coverage of 90\%. Figure~\ref{fig:examplesCI} shows the predictions for three profiles with different $\Xmax$ values, using networks trained on the three detector-level datasets described in Section \ref{sec:training}.

The third term compares the profile reconstructed by the physics branch with the input profile at measured points only, using the mean squared error. The two physics terms play complementary roles: $\mathcal{L}_\mathrm{GH}$ anchors the predicted parameters to the fit, while $\mathcal{L}_\mathrm{rec}$ requires them to reproduce the measured profile itself. Both terms depend only on the measured profile and on a fit to it. Of the three terms, only $\mathcal{L}_q$ requires the simulated $\Xfirst$, so the physics terms do not rely on simulated quantities.

\subsection{Training}
\label{sec:training}

We trained the network on an NVIDIA GeForce RTX 4060 Ti GPU with a batch size of 128 for up to 300 epochs, requiring less than one minute per epoch. This short training time allowed us to train separate networks for each dataset in Section~\ref{sec:data} (EPOS LHC-R, \textsc{Sibyll} 2.3d, and Mix), for both detector-level and ideal profiles. The training subsets contain $2.8 \times 10^{6}$ events for Mix and $1.4 \times 10^{6}$ events for each single-model dataset. We used the Adam optimizer~\cite{kingma2017}, with a learning rate linearly increased during the first epoch to $2 \times 10^{-3}$. The learning rate was subsequently reduced by a factor of 0.3 whenever the validation mean absolute error (MAE) of $\Xfirst$ failed to improve for 10 epochs, down to a minimum of $10^{-6}$. Since the pinball loss at $\tau = 0.5$ is proportional to the absolute error, the MAE is the natural metric for the median estimate.

The loss weights in Eq.~\eqref{eq:loss} followed a schedule to improve convergence. During the first 30 epochs, $(w_\mathrm{GH}, w_q, w_\mathrm{rec}) = (1.0, 0.5, 0.1)$. They were then changed to $(0.1, 1.0, 0.8)$, shifting the emphasis toward $X_\mathrm{first}$ while retaining the reconstruction term~\cite{bengio2009}. Training was stopped if the validation MAE, expressed in $\gcm$, failed to improve for 30 epochs after the transition, and the weights from the best epoch were retained.

\section{Results: Event-by-event reconstruction}
\label{sec:res1}

\subsection{Information loss in the detection process}

\begin{figure}[htbp]
    \centering
    \includegraphics[width=0.85\linewidth]{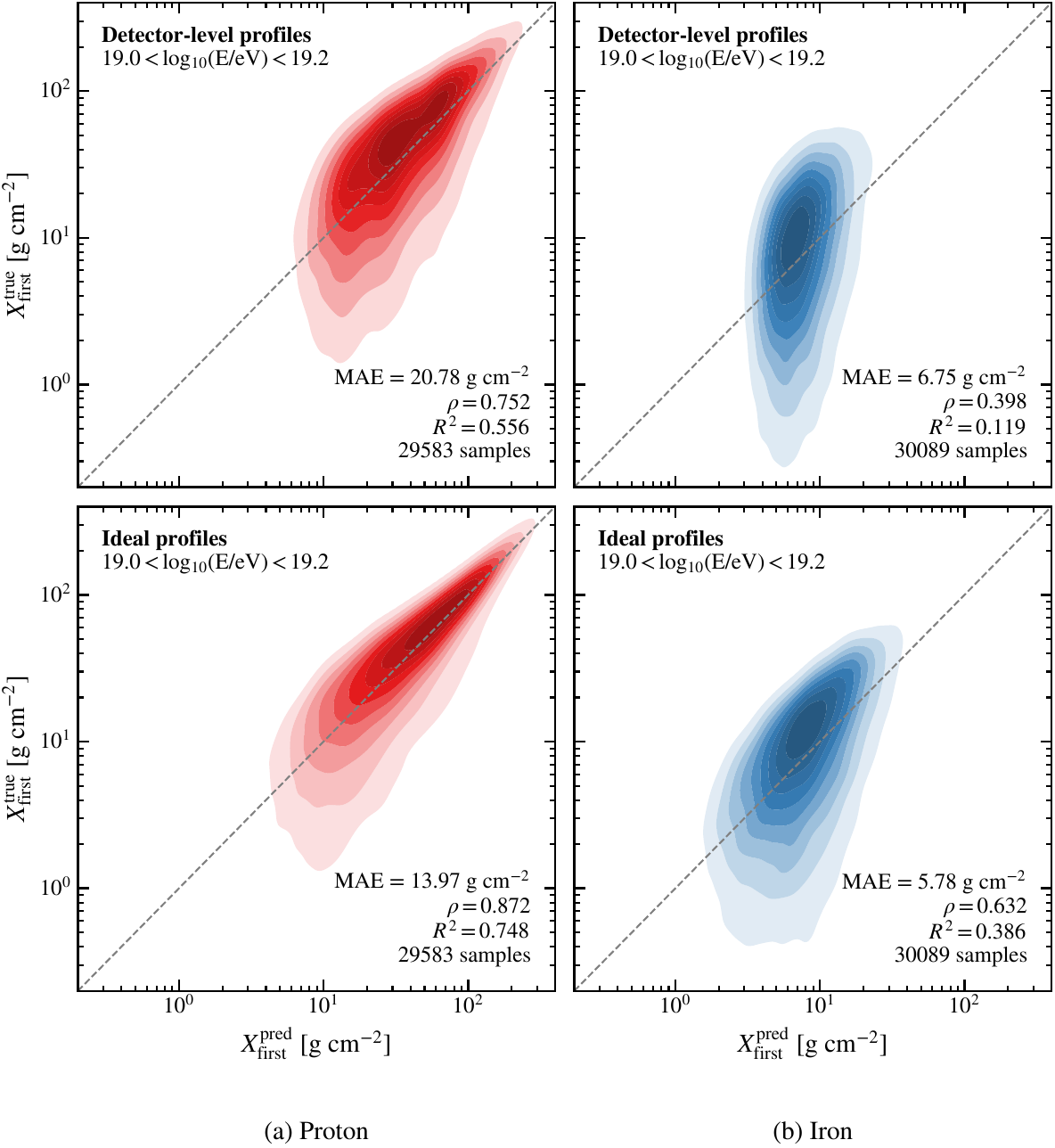}
    \caption{True versus predicted $\Xfirst$ for (a) protons and (b) iron nuclei in the energy bin $19.0 \leq \lgE < 19.2$. Results are shown for networks trained on detector-level (top) and ideal (bottom) profiles. Both networks were trained and evaluated on the Mix dataset, containing events simulated with \Epos{} and \Sibyll{} for both primary types.}
    \label{fig:ideal_vs_det}
\end{figure}

We first assess how detection affects the information on $\Xfirst$ contained in the longitudinal profiles by comparing two networks with identical architectures, trained on ideal and detector-level profiles. Both networks were trained and evaluated on the Mix dataset, which includes both hadronic interaction models and both primaries, and their performance is shown separately for each primary. Figure~\ref{fig:ideal_vs_det} shows the true versus predicted $\Xfirst$ in the representative energy bin of $10^{19.0-19.2}$~eV. The mean absolute error is reported in each plot alongside the Pearson correlation coefficient ($\rho$) and the coefficient of determination ($R^2$).

For protons, the ideal profiles yield $\rho = 0.87$ and ${\rm MAE} \approx 14~\gcm$. With the detected profiles, the correlation decreases to $\rho = 0.75$ and the absolute error increases to ${\rm MAE} \approx 21~\gcm$. Despite this degradation, the results indicate that event-by-event reconstruction of $\Xfirst$ remains feasible under realistic detection conditions.

For iron, reconstruction is already more challenging at the ideal-profile level, with a correlation of $\rho = 0.63$, which drops to $\rho = 0.40$ with the detected profiles. The mean absolute errors are comparable to the iron interaction mean free path, $\lambda_{\mathrm{Fe}} \approx 10~\gcm$, highlighting the difficulty of resolving fluctuations in the depth of the first interaction. The degradation due to detector effects is consistent with Fig.~\ref{fig:cuts} (right): because iron showers develop earlier, many selected events only marginally satisfy the requirement that at least $200~\gcm$ of the profile be observed before $\Xmax$. Consequently, much of the rising edge, which contains important information about $\Xfirst$, remains unobserved.

\begin{figure}[htbp]
    \centering
    \includegraphics[width=1\linewidth]{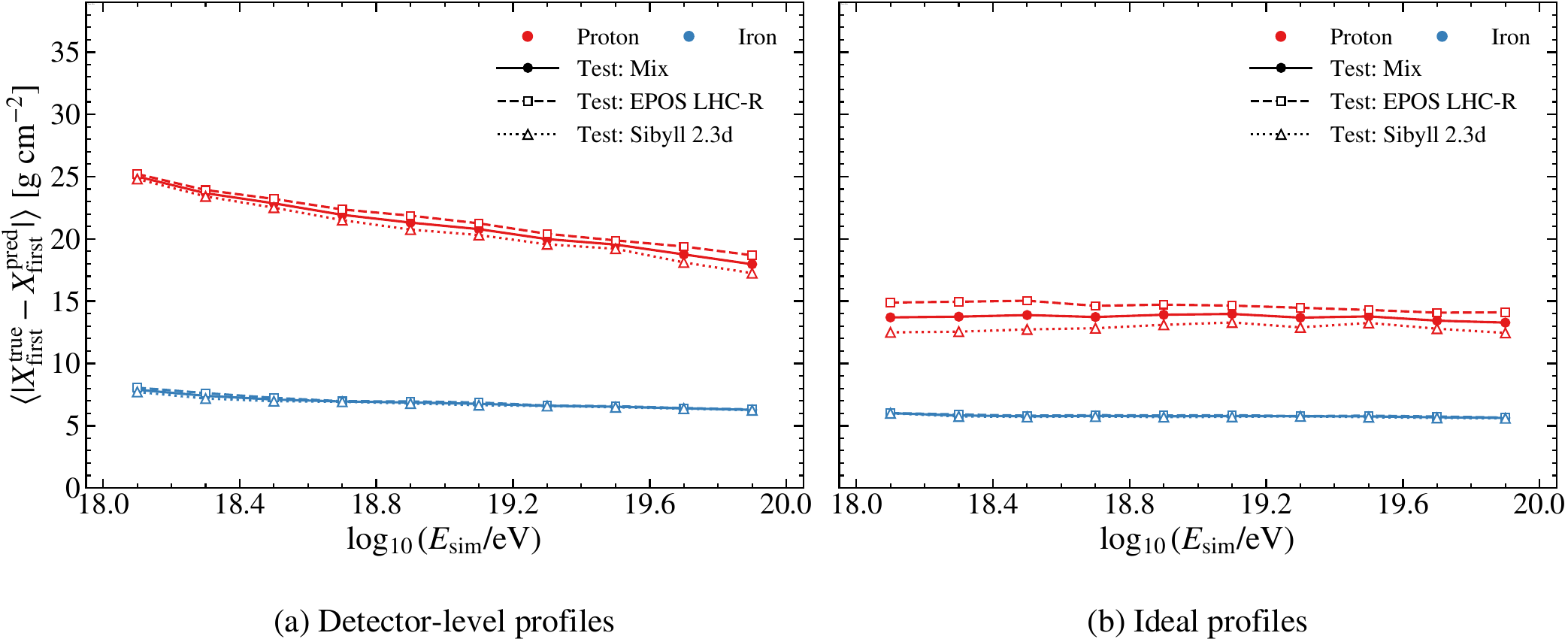}
    \caption{Mean absolute error as a function of primary-particle energy for neural networks trained and evaluated on (a) detector-level longitudinal profiles and (b) ideal profiles. Both networks were trained on the Mix dataset, containing events simulated with \Epos{} and \Sibyll{} for both primary types.}
    \label{fig:mae_vs_en}
\end{figure}

The performance of our method improves with increasing primary-particle energy. The degradation observed when using detector-level profiles, compared with ideal profiles, is consistent across all trained networks, regardless of the hadronic interaction model used for evaluation. In all cases, the reconstruction errors and correlation coefficients show only minor variations between hadronic interaction models. Figure~\ref{fig:mae_vs_en} illustrates this behavior for the network trained on the Mix dataset. In the following, we focus on the performance of models trained and evaluated using detector-level profiles.

\subsection{Model performance on detector-level profiles}

\begin{figure}[htbp]
    \centering
    \includegraphics[width=1\linewidth]{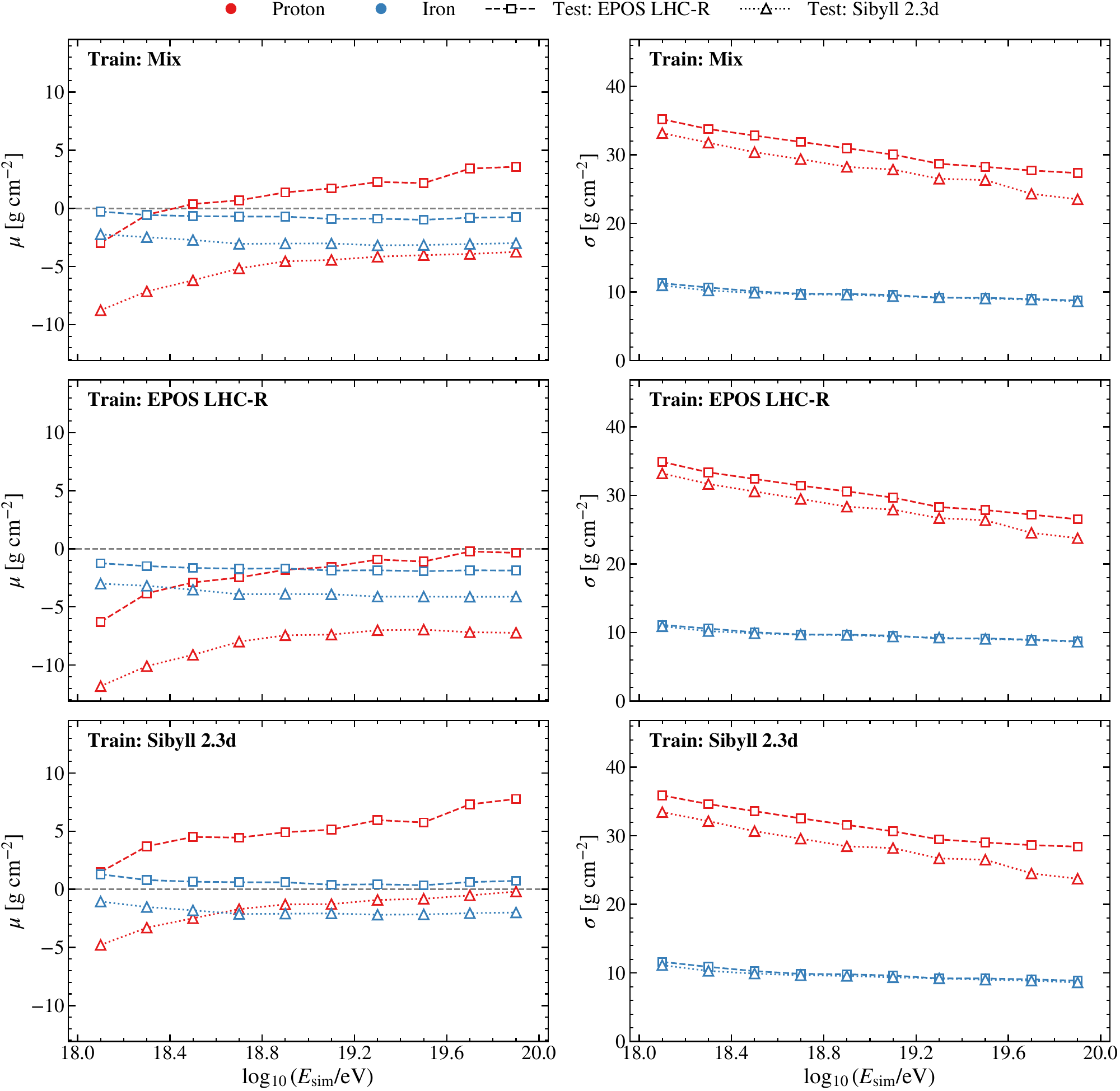}
    \caption{Energy-dependent bias (left) and resolution (right) of the $\Xfirst$ predictions of each neural network for protons and iron across the different hadronic models used in the training and evaluation of the networks.}
    \label{fig:bias_res}
\end{figure}

In addition to the MAE, we also evaluate the mean ($\mu$) and standard deviation ($\sigma$) of the residuals, $\Xfirst^\mathrm{true}-\Xfirst^\mathrm{pred}$, which quantify the bias and resolution, respectively. Since the point estimate is the conditional median, $\mu$ is not expected to vanish exactly even for an ideal estimator when the conditional distribution is asymmetric. Figure~\ref{fig:bias_res} shows these metrics as functions of primary energy for different combinations of hadronic interaction models used in training and evaluation.

The resolution remains below $40~\gcm$ for protons and is approximately $10~\gcm$ for iron. For protons, the resolution improves with increasing energy and depends on the hadronic model used to generate the profiles: Sibyll-generated profiles yield consistently smaller values than EPOS-generated profiles, regardless of the training model. For iron, the resolution is nearly constant with energy and shows little dependence on the hadronic model.

The most prominent differences appear in the bias, which depends on the combination of models used for training and evaluation. Training on the combined Mix dataset generally reduces the model-dependent bias compared with training on either hadronic model alone. Nevertheless, for networks trained on one model and evaluated on the other, the absolute bias generally remains below approximately $10~\gcm$ at energies above $10^{18.2}~\mathrm{eV}$.

To assess the calibration of the predicted uncertainties, we examined the empirical coverage of the interval $[\hat q_{0.05}, \hat q_{0.95}]$, i.e.\ the fraction of events whose true $\Xfirst$ falls inside it (Table~\ref{tab:coverage}). When the network is trained and tested with the same hadronic interaction model, the coverage lies between 89.4\% and 90.0\% for both primaries, in agreement with the nominal 90\%, and is stable with energy. In the cross-model tests, the misses become asymmetric. The network trained on \Epos{} undercovers the \Sibyll{} events (86.7\% for protons and 85.2\% for iron), with the true $\Xfirst$ above $\hat q_{0.95}$ in about 10\% of the cases. In the opposite direction, the coverage remains close to nominal, but the misses concentrate below $\hat q_{0.05}$. The network trained on the Mix dataset reaches 88.7--91.2\% in both test datasets. The direction of these deviations agrees with the model-dependent bias of the interaction length discussed in Section~\ref{sec:res2}.

\begin{table}[htbp]
  \centering
  \caption{Empirical coverage (\%) of the interval $[\hat q_{0.05}, \hat q_{0.95}]$,
    whose nominal coverage is 90\%, for each combination of training and test datasets,
    integrated over energy, together with the fractions of events with the true $\Xfirst$
    below $\hat q_{0.05}$ and above $\hat q_{0.95}$ (nominal 5\% each). The statistical
    uncertainties are about 0.1\%.}
  \label{tab:coverage}
  \small
  \begin{tabular}{ll ccc ccc}
    \toprule
    & & \multicolumn{3}{c}{Proton} & \multicolumn{3}{c}{Iron} \\
    \cmidrule(lr){3-5} \cmidrule(lr){6-8}
    Train & Test & Cov. & Below & Above & Cov. & Below & Above \\
    \midrule
    EPOS LHC-R           & EPOS LHC-R           & 89.4 & 4.8 & 5.8 & 89.7 & 5.6 & 4.7 \\
    \textsc{Sibyll} 2.3d & \textsc{Sibyll} 2.3d & 90.0 & 4.8 & 5.2 & 89.8 & 5.4 & 4.9 \\
    \midrule
    Mix                  & EPOS LHC-R           & 90.5 & 5.5 & 4.0 & 91.2 & 6.0 & 2.8 \\
    Mix                  & \textsc{Sibyll} 2.3d & 88.9 & 4.1 & 7.0 & 88.7 & 4.7 & 6.6 \\
    \midrule
    EPOS LHC-R           & \textsc{Sibyll} 2.3d & 86.7 & 3.6 & 9.7 & 85.2 & 4.2 & 10.6 \\
    \textsc{Sibyll} 2.3d & EPOS LHC-R           & 90.7 & 6.6 & 2.8 & 91.2 & 7.0 & 1.9 \\
    \bottomrule
  \end{tabular}
\end{table}

\begin{figure}[htbp]
    \centering
    \includegraphics[width=1\linewidth]{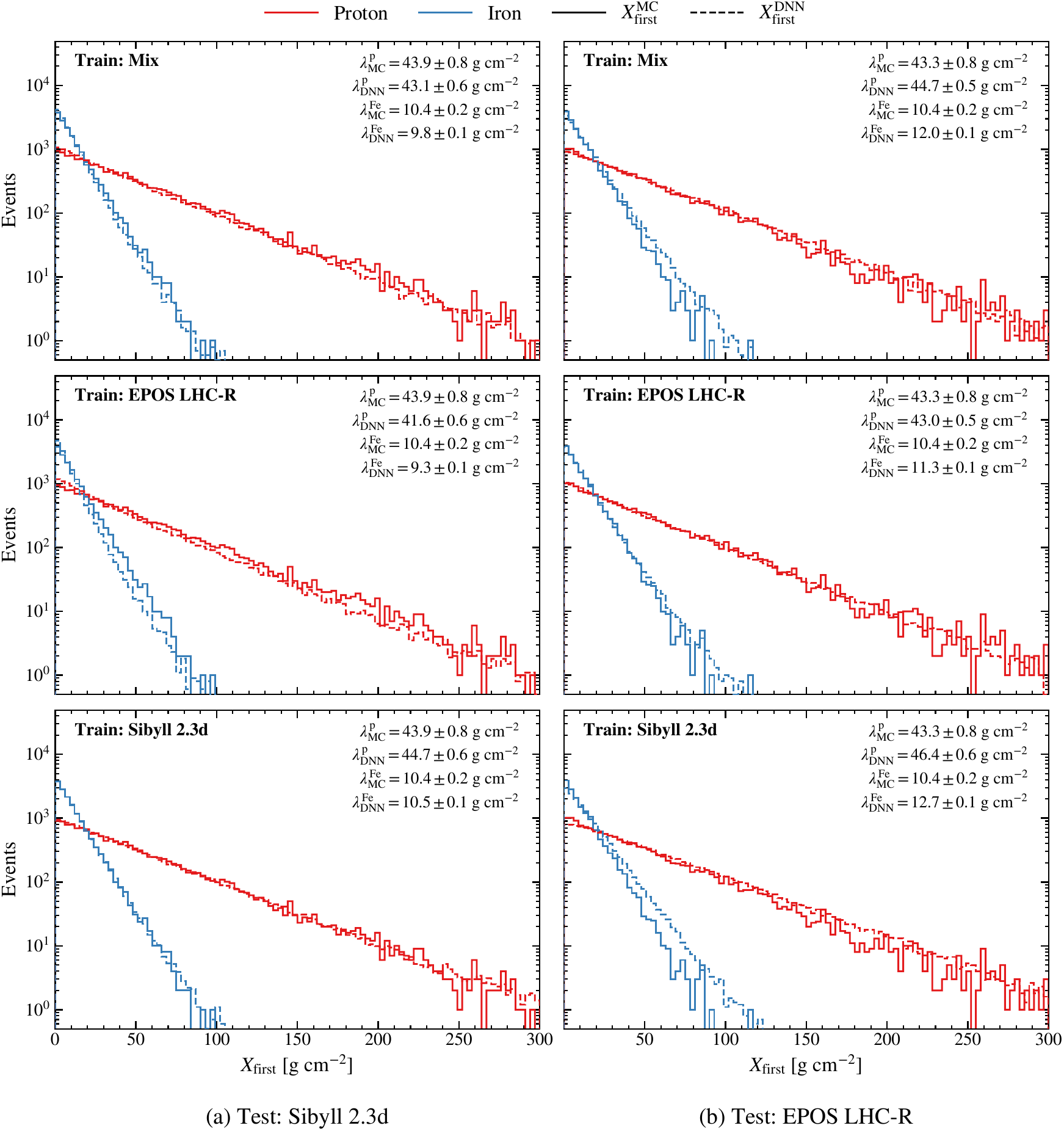}
    \caption{True $\Xfirst$ distributions from the datasets (MC) and from the predictions of each neural network (DNN) for protons and iron across different hadronic models used in the training and evaluation of the networks. The energy bin is $19.0 \leq \lgE < 19.2$.}
    \label{fig:dist_rec}
\end{figure}

We also assess whether the method reproduces the full $\Xfirst$ distributions, beyond their first moments. Figure~\ref{fig:dist_rec} compares reference and reconstructed distributions for protons and iron in the energy bin of $10^{19.0-19.2}$~eV, using different combinations of hadronic interaction models for training and evaluation. Each predicted distribution was built by sampling, for every event, the conditional distribution defined by the seven quantiles predicted by each neural network, while the true distributions were built from the true $\Xfirst$ values of each dataset. Each panel reports the mean interaction depths, $\lambda_{\mathrm{MC}}$ and $\lambda_{\mathrm{DNN}}$, obtained from exponential fits to each distribution in dataset and network predictions, respectively. 

The results show a clear dependence on the hadronic interaction model. Distributions from networks trained on one model and evaluated on profiles generated with the other can overestimate or underestimate the mean interaction depth. This is particularly pronounced for networks trained on \Sibyll{} and evaluated on \Epos{} profiles, where the reconstructed mean interaction depths exceed the reference values for both protons and iron. These discrepancies indicate that, although the network reproduces important features of the $\Xfirst$ distributions, differences between hadronic models can introduce systematic shifts in the reconstructed distributions.

\section{Results: Proton--air interaction length reconstruction}
\label{sec:res2}

\begin{figure}[htbp]
    \centering
    \includegraphics[width=0.8\linewidth]{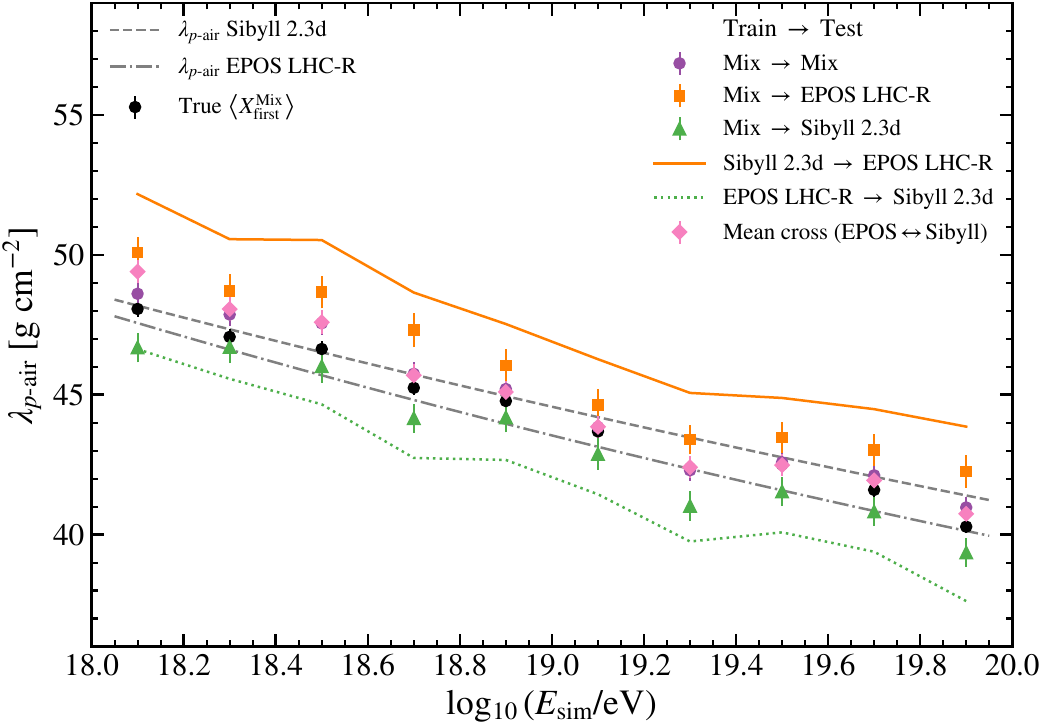}
    \caption{Proton--air interaction length as a function of energy. The gray lines show $\lambda_{\mathrm{p\text{-}air}}$ of \Sibyll{} (dashed) and \Epos{} (dash-dotted) from \Conex{}, and the black circles the mean true $\Xfirst$ of the protons in the Mix test set. Colored markers show the interaction lengths reconstructed by the Mix network on the Mix (circles), \Epos{} (squares), and \Sibyll{} (triangles) test sets. The lines show the cross-model tests, \Sibyll{} $\rightarrow$ \Epos{} (solid) and \Epos{} $\rightarrow$ \Sibyll{} (dotted), and the diamonds their average. The reconstructed values were obtained from an exponential fit to the 20\% most deeply penetrating events of the reconstructed $\Xfirst$ distribution, in bins of $\Delta\log_{10}(E/\mathrm{eV}) = 0.2$.}
    \label{fig:lambda}
\end{figure}

Figure~\ref{fig:dist_rec} showed the $\Xfirst$ distributions of each primary in a single energy bin, from which the interaction lengths follow directly as $\lambda_{A\text{-air}} = \langle\Xfirst\rangle$. In this section, we extend this analysis to proton showers over the full energy range of this study. Similarly to the analyses of the $X_\mathrm{max}$ distribution~\cite{TAcross2015,PAcross2012}, the proton--air interaction length was estimated from an exponential fit to the tail of the reconstructed $\Xfirst$ distribution, defined by the 20\% most deeply penetrating events, in energy bins of $\Delta\log_{10}(E/\mathrm{eV}) = 0.2$. Since the fit is applied to $\Xfirst$ itself, no conversion factor between the fitted slope and  $\lambda_{\mathrm{p\text{-}air}}$ is required. The results are shown in Figure~\ref{fig:lambda}.

Applied to the true $\Xfirst$ values, the same procedure reproduces the interaction length of \Conex{} within the statistical uncertainties, with average deviations below 0.2~$\gcm$ ($<0.4\%$), so the estimator itself introduces no measurable bias.  When training and test share the same hadronic interaction model (not shown in Figure~\ref{fig:lambda} for clarity), the reconstructed interaction length agrees with the true value within 0.2~$\gcm$ for \Sibyll{} and 0.8~$\gcm$ ($1.8\%$) for \Epos{}, the latter mainly at the lowest energies. In the cross-model tests, the bias reaches $+3.7~\gcm$ ($+8.4\%$) for the network trained on \Sibyll{} and applied to \Epos{}, and $-2.7~\gcm$ ($-6.0\%$) in the opposite case. Since the tail slope is insensitive to a global shift of the predictions, these deviations reflect a different scale of the reconstructed distributions. 

The network trained on the Mix dataset lies between these cases, with nearly energy-independent biases of $+2.0~\gcm$ ($+4.7\%$) on \Epos{} and $-1.4~\gcm$ ($-3.1\%$) on \Sibyll{}. Accordingly, the average of the two cross-model results agrees with that of the Mix network, which therefore  behaves as an effective model between the two. Training on both models reduces but does not remove the model dependence, and half the difference between the two tests, about 4\% on $\lambda_{\mathrm{p\text{-}air}}$, can be taken as a systematic uncertainty. 

The result is robust against the presence of heavier primaries. Since protons penetrate deeper on average, the tail of the $\Xfirst$ distribution is dominated by them: increasing the iron fraction of the test sample up to 60\% changes the reconstructed $\lambda_{\mathrm{p\text{-}air}}$ by less than 1\%, although the network was trained with equal numbers of proton and iron showers. Intermediate primaries, in particular helium, are expected to have a stronger impact, since their $\Xfirst$ distribution overlaps more with the proton tail. The fraction of 20\% follows the choice of Ref.~\cite{PAcross2012}, where it was shown to keep the bias from helium fractions up to 25\% below the statistical uncertainty of the $X_\mathrm{max}$ analysis. Helium was not included in our simulations, and its effect on the $\Xfirst$ tail remains to be quantified.

The datasets used in this work were constructed so that the event selection does not distort the $\Xfirst$ distributions (Section~\ref{sec:data}). In measured data, however, each shower is observed with a single geometry, and showers with deeper first interactions are more likely to fail the trigger or have their $\Xmax$ outside the field of view. To assess this effect, we weighted the test events by the selection efficiency obtained from simulations without redrawing geometries for each shower, thereby reproducing the $\Xfirst$ distribution of the selected events. The interaction length inferred by the Mix network above $10^{18.6}$~eV decreases by about 12\% under this reweighting, while the model dependence remains unchanged. Fiducial cuts on the shower geometry, as used in the hybrid analyses of the Pierre Auger Observatory~\cite{abreuExposureHybridDetector2011,PAcross2012}, largely suppress this effect. Requiring the field of view to cover slant depths beyond 1000~$\gcm$ and the impact parameter to be below 15~km reduces the bias to about 1\% without increasing the model-related systematic uncertainty, while reducing the available event sample by a factor of 2--4. At lower energies, energy-dependent cuts, including a requirement on the upper edge of the field of view, would be necessary. Such fiducial cuts must therefore be included in an application to data.

\section{Conclusions}
\label{sec:conc}

We have presented a physics-motivated neural network that reconstructs the depth of the first interaction, $\Xfirst$, of individual air showers from their longitudinal profiles measured by a fluorescence detector. The network combines a temporal convolutional encoder with two branches: a decoder that implements the Gaisser--Hillas function, which ties part of the latent representation to physical parameters of the profile, and a quantile branch that estimates the conditional distribution of $\Xfirst$ for each event. The method was trained and evaluated on proton- and iron-induced showers between $10^{18}$ and $10^{20}$~eV, simulated with \Epos{} and \Sibyll{} and processed by an end-to-end simulation of the detector response with quality cuts similar to those applied to real data.

Comparing networks trained on ideal and detector-level profiles quantifies the information lost in the detection process. For protons, the correlation between reconstructed and true $\Xfirst$ decreases from 0.87 to 0.75 in the energy bin $19.0 \leq \lgE < 19.2$, and the mean absolute error of the detector-level reconstruction decreases from about 25 to 18~\gcm\ over the full energy range, whereas it is nearly constant for ideal profiles. For iron, the information on $\Xfirst$ is limited already at the ideal level and is largely lost after detection, since iron showers develop early and the rising edge of their profiles is mostly unobserved. The resolution depends only weakly on the hadronic interaction model, while the bias depends on the combination of models used for training and evaluation and remains below about 10~\gcm\ above $10^{18.2}$~eV in the cross-model tests.

The predicted distributions reproduce the exponential shape of the true $\Xfirst$ distributions. This allows the proton--air interaction length to be estimated from an exponential fit to their tail, without the simulation-based conversion required in analyses of the $\Xmax$ distribution. When training and test share the same hadronic model, the interaction length is recovered within 2\%. The network trained on both models yields biases of $+4.7\%$ and $-3.1\%$ on \Epos{} and \Sibyll{}, respectively, and behaves as an effective model between the two. Half of this difference, about 4\%, provides an estimate of the model-related systematic uncertainty. The result changes by less than 1\% for iron fractions of up to 60\% in the test sample.

These results show that the event-by-event reconstruction of $\Xfirst$ is feasible for protons under realistic detection conditions and that the reconstructed distributions offer a complementary route to measurements of the proton--air cross section. Several aspects must be addressed before an application to data. 
The selection efficiency depends on $\Xfirst$ and must be controlled, for example with fiducial cuts on the geometry, which we have shown to reduce its effect on the reconstructed interaction length.
The model dependence was assessed with only two hadronic interaction models and should be extended to others. Intermediate primaries, in particular helium, were not simulated, and their effect on the tail of the $\Xfirst$ distribution remains to be quantified. Despite these limitations, since the physics terms of the loss do not rely on simulated quantities, they could be used to adapt the network to measured profiles, reducing its dependence on simulations. 

\acknowledgments

This study was financed, in part, by the S\~{a}o Paulo Research Foundation (FAPESP), Brazil. Process Numbers 2021/01089-1 and 2024/08946-5. We thank the National Laboratory for Scientific Computing (LNCC/MCTI,  Brazil) for providing HPC resources on the SDumont supercomputer (http://sdumont.lncc.br).

\section*{Data availability}

Data supporting this study are available from the corresponding author upon reasonable request.

\bibliographystyle{JHEP}
\bibliography{biblio.bib}

\providecommand{\href}[2]{#2}\begingroup\raggedright\begin{thebibliography}{10}

\bibitem{kampert2012}
K.-H.~Kampert and M.~Unger, \emph{Measurements of the cosmic ray composition
  with air shower experiments},
  \href{https://doi.org/10.1016/j.astropartphys.2012.02.004}{\emph{Astropart.
  Phys.} {\bfseries 35} (2012) 660}
  [\href{https://arxiv.org/abs/1201.0018}{{\ttfamily 1201.0018}}].

\bibitem{souza2017}
V.~de~Souza, \emph{Measurements of the depth of maximum of air-shower profiles
  at the {Pierre Auger Observatory} and their composition implications},
  \href{https://arxiv.org/abs/1701.06812}{{\ttfamily 1701.06812}}.

\bibitem{bellido_2017}
J.~Bellido, \emph{Depth of maximum of air-shower profiles at the {Pierre Auger
  Observatory}: measurements above $10^{17.2}$ {eV} and composition
  implications},  in \emph{35th International Cosmic Ray Conference
  (ICRC2017)}, vol.~ICRC2017, p.~506, 2017,
  \href{https://pos.sissa.it/301/506}{https://pos.sissa.it/301/506}.

\bibitem{batista_2019}
R.~Alves~Batista, J.~Biteau, M.~Bustamante, K.~Dolag, R.~Engel, K.~Fang et~al.,
  \emph{Open questions in cosmic-ray research at ultrahigh energies},
  \href{https://doi.org/10.3389/fspas.2019.00023}{\emph{Front. Astron. Space
  Sci.} {\bfseries 6} (2019) 23}
  [\href{https://arxiv.org/abs/1903.06714}{{\ttfamily 1903.06714}}].

\bibitem{PAcross2012}
{\scshape Pierre Auger} collaboration, \emph{Measurement of the proton-air
  cross section at $\sqrt{s}=57$ {TeV} with the {Pierre Auger Observatory}},
  \href{https://doi.org/10.1103/PhysRevLett.109.062002}{\emph{Phys. Rev. Lett.}
  {\bfseries 109} (2012) 062002}
  [\href{https://arxiv.org/abs/1208.1520}{{\ttfamily 1208.1520}}].

\bibitem{TAcross2015}
{\scshape Telescope Array} collaboration, \emph{Measurement of the proton-air
  cross section with {Telescope Array}'s {Middle Drum} detector and surface
  array in hybrid mode},
  \href{https://doi.org/10.1103/PhysRevD.92.032007}{\emph{Phys. Rev. D}
  {\bfseries 92} (2015) 032007}
  [\href{https://arxiv.org/abs/1505.01860}{{\ttfamily 1505.01860}}].

\bibitem{ulrichMeasurementProtonairCross2009}
R.~Ulrich, J.~Bl{\"u}mer, R.~Engel, F.~Sch{\"u}ssler and M.~Unger, \emph{On the
  measurement of the proton-air cross section using air shower data},
  \href{https://doi.org/10.1088/1367-2630/11/6/065018}{\emph{New J. Phys.}
  {\bfseries 11} (2009) 065018}
  [\href{https://arxiv.org/abs/0903.0404}{{\ttfamily 0903.0404}}].

\bibitem{andringaMassCompositionCrosssection2011}
S.~Andringa, R.~Concei{\c c}{\~a}o and M.~Pimenta, \emph{Mass composition and
  cross-section from the shape of cosmic ray shower longitudinal profiles},
  \href{https://doi.org/10.1016/j.astropartphys.2010.10.002}{\emph{Astropart.
  Phys.} {\bfseries 34} (2011) 360}.

\bibitem{yapici_2015}
T.~Yapici, \emph{A method for determining the mass composition of ultra-high
  energy cosmic rays by predicting the depth of first interaction of individual
  extensive air showers}, Ph.D. thesis, Michigan Technological University,
  Houghton, Michigan, 2015.
\newblock
  \href{https://digitalcommons.mtu.edu/etds/1019/}{https://digitalcommons.mtu.edu/etds/1019/}.

\bibitem{abdulhalim_2025}
{\scshape Pierre Auger} collaboration, \emph{Inference of the mass composition
  of cosmic rays with energies from $10^{18.5}$ to $10^{20}$ {eV} using the
  {Pierre Auger Observatory} and deep learning},
  \href{https://doi.org/10.1103/PhysRevLett.134.021001}{\emph{Phys. Rev. Lett.}
  {\bfseries 134} (2025) 021001}
  [\href{https://arxiv.org/abs/2406.06315}{{\ttfamily 2406.06315}}].

\bibitem{Auger2021XmaxDNN}
{\scshape Pierre Auger} collaboration, \emph{Deep-learning based reconstruction
  of the shower maximum {X$_{\mathrm{max}}$} using the water-{Cherenkov}
  detectors of the {Pierre Auger Observatory}},
  \href{https://doi.org/10.1088/1748-0221/16/07/P07019}{\emph{JINST} {\bfseries
  16} (2021) P07019} [\href{https://arxiv.org/abs/2101.02946}{{\ttfamily
  2101.02946}}].

\bibitem{wang_2026}
Z.~Wang, E.~Mayotte, S.~Mayotte, N.~Woo, J.~Burton-Heibges, N.~San~Martin
  et~al., \emph{Prospects for deep-learning-based mass reconstruction of
  ultra-high-energy cosmic rays using simulated air-shower profiles},
  \href{https://doi.org/10.1088/1475-7516/2026/06/069}{\emph{JCAP} {\bfseries
  06} (2026) 069} [\href{https://arxiv.org/abs/2508.13933}{{\ttfamily
  2508.13933}}].

\bibitem{kashinath2021}
K.~Kashinath, M.~Mustafa, A.~Albert, J.-L.~Wu, C.~Jiang, S.~Esmaeilzadeh
  et~al., \emph{Physics-informed machine learning: case studies for weather and
  climate modelling}, \href{https://doi.org/10.1098/rsta.2020.0093}{\emph{Phil.
  Trans. Roy. Soc. A} {\bfseries 379} (2021) 20200093}.

\bibitem{yang2022}
T.-Y.~Yang, J.~Rosca, K.~Narasimhan and P.J.~Ramadge, \emph{Learning physics
  constrained dynamics using autoencoders},  in \emph{Advances in Neural
  Information Processing Systems 35 (NeurIPS 2022)}, pp.~17157--17172, 2022,
  \href{https://doi.org/10.52202/068431-1248}{https://doi.org/10.52202/068431-1248}.

\bibitem{conex1}
T.~Pierog, M.K.~Alekseeva, T.~Bergmann, V.~Chernatkin, R.~Engel, D.~Heck
  et~al., \emph{First results of fast one-dimensional hybrid simulation of
  {EAS} using {CONEX}},
  \href{https://doi.org/10.1016/j.nuclphysbps.2005.07.029}{\emph{Nucl. Phys. B
  Proc. Suppl.} {\bfseries 151} (2006) 159}
  [\href{https://arxiv.org/abs/astro-ph/0411260}{{\ttfamily
  astro-ph/0411260}}].

\bibitem{conex2}
T.~Bergmann, R.~Engel, D.~Heck, N.N.~Kalmykov, S.~Ostapchenko, T.~Pierog
  et~al., \emph{One-dimensional hybrid approach to extensive air shower
  simulation},
  \href{https://doi.org/10.1016/j.astropartphys.2006.08.005}{\emph{Astropart.
  Phys.} {\bfseries 26} (2007) 420}
  [\href{https://arxiv.org/abs/astro-ph/0606564}{{\ttfamily
  astro-ph/0606564}}].

\bibitem{pierogEPOSLHCRGlobalApproach2025}
T.~Pierog and K.~Werner, \emph{{EPOS.LHC-R}: a global approach to solve the
  muon puzzle},  in \emph{39th International Cosmic Ray Conference (ICRC2025)},
  vol.~ICRC2025, p.~358, 2025,
  \href{https://pos.sissa.it/501/358}{https://pos.sissa.it/501/358}
  [\href{https://arxiv.org/abs/2508.07105}{{\ttfamily 2508.07105}}].

\bibitem{SIBYLL2020}
F.~Riehn, R.~Engel, A.~Fedynitch, T.K.~Gaisser and T.~Stanev, \emph{Hadronic
  interaction model {Sibyll} 2.3d and extensive air showers},
  \href{https://doi.org/10.1103/PhysRevD.102.063002}{\emph{Phys. Rev. D}
  {\bfseries 102} (2020) 063002}
  [\href{https://arxiv.org/abs/1912.03300}{{\ttfamily 1912.03300}}].

\bibitem{arapy_zenodo}
M.~Olegario and V.~de~Souza, ``{ARAPy} v1.0.0 [software].'' Zenodo,
  \href{https://doi.org/10.5281/zenodo.23705631}{doi:10.5281/zenodo.23705631},
  Oct., 2026.

\bibitem{abreuExposureHybridDetector2011}
{\scshape Pierre Auger} collaboration, \emph{The exposure of the hybrid
  detector of the {Pierre Auger Observatory}},
  \href{https://doi.org/10.1016/j.astropartphys.2010.10.001}{\emph{Astropart.
  Phys.} {\bfseries 34} (2011) 368}
  [\href{https://arxiv.org/abs/1010.6162}{{\ttfamily 1010.6162}}].

\bibitem{abrahamFDAuger2010}
{\scshape Pierre Auger} collaboration, \emph{The fluorescence detector of the
  {Pierre Auger Observatory}},
  \href{https://doi.org/10.1016/j.nima.2010.04.023}{\emph{Nucl. Instrum. Meth.
  A} {\bfseries 620} (2010) 227}
  [\href{https://arxiv.org/abs/0907.4282}{{\ttfamily 0907.4282}}].

\bibitem{carvalhoEffectsEnergyError2007}
W.~Carvalho, I.F.M.~Albuquerque and V.~de~Souza, \emph{Effects of the energy
  error distribution of fluorescence telescopes on the {UHECR} energy
  spectrum},
  \href{https://doi.org/10.1016/j.astropartphys.2007.04.010}{\emph{Astropart.
  Phys.} {\bfseries 28} (2007) 89}.

\bibitem{thepierreaugercollaborationMeasurementAverageShape2018}
{\scshape Pierre Auger} collaboration, \emph{Measurement of the average shape
  of longitudinal profiles of cosmic-ray air showers at the {Pierre Auger
  Observatory}},
  \href{https://doi.org/10.1088/1475-7516/2019/03/018}{\emph{JCAP} {\bfseries
  03} (2019) 018} [\href{https://arxiv.org/abs/1811.04660}{{\ttfamily
  1811.04660}}].

\bibitem{gaisserhillasref}
T.K.~Gaisser and A.M.~Hillas, \emph{Reliability of the method of constant
  intensity cuts for reconstructing the average development of vertical
  showers},  in \emph{Proceedings of the 15th International Cosmic Ray
  Conference}, vol.~8, (Plovdiv, Bulgaria), p.~353, 1977,
  \href{https://articles.adsabs.harvard.edu/pdf/1977ICRC...15h.353G}{https://articles.adsabs.harvard.edu/pdf/1977ICRC...15h.353G}.

\bibitem{bai2018}
S.~Bai, J.Z.~Kolter and V.~Koltun, \emph{An empirical evaluation of generic
  convolutional and recurrent networks for sequence modeling},
  \href{https://arxiv.org/abs/1803.01271}{{\ttfamily 1803.01271}}.

\bibitem{he2015}
K.~He, X.~Zhang, S.~Ren and J.~Sun, \emph{Deep residual learning for image
  recognition},  in \emph{2016 IEEE Conference on Computer Vision and Pattern
  Recognition (CVPR)}, pp.~770--778, 2016,
  \href{https://doi.org/10.1109/CVPR.2016.90}{https://doi.org/10.1109/CVPR.2016.90}
  [\href{https://arxiv.org/abs/1512.03385}{{\ttfamily 1512.03385}}].

\bibitem{ba2016}
J.L.~Ba, J.R.~Kiros and G.E.~Hinton, \emph{Layer normalization},
  \href{https://arxiv.org/abs/1607.06450}{{\ttfamily 1607.06450}}.

\bibitem{chernozhukov2010}
V.~Chernozhukov, I.~Fern{\'a}ndez-Val and A.~Galichon, \emph{Quantile and
  probability curves without crossing},
  \href{https://doi.org/10.3982/ECTA7880}{\emph{Econometrica} {\bfseries 78}
  (2010) 1093} [\href{https://arxiv.org/abs/0704.3649}{{\ttfamily 0704.3649}}].

\bibitem{cannon2018}
A.J.~Cannon, \emph{Non-crossing nonlinear regression quantiles by monotone
  composite quantile regression neural network, with application to rainfall
  extremes}, \href{https://doi.org/10.1007/s00477-018-1573-6}{\emph{Stoch.
  Environ. Res. Risk Assess.} {\bfseries 32} (2018) 3207}.

\bibitem{tensorflow2015}
M.~Abadi, A.~Agarwal, P.~Barham, E.~Brevdo, Z.~Chen, C.~Citro et~al.,
  ``{TensorFlow}: large-scale machine learning on heterogeneous systems.''
  \url{https://www.tensorflow.org/}, 2015.

\bibitem{chollet2015keras}
F.~Chollet et~al., ``Keras.'' \url{https://keras.io}, 2015.

\bibitem{bergstra2012}
J.~Bergstra and Y.~Bengio, \emph{Random search for hyper-parameter
  optimization}, {\emph{J. Mach. Learn. Res.} {\bfseries 13} (2012) 281}.

\bibitem{koenker1978}
R.~Koenker and G.~Bassett, \emph{Regression quantiles},
  \href{https://doi.org/10.2307/1913643}{\emph{Econometrica} {\bfseries 46}
  (1978) 33}.

\bibitem{koenker2005}
R.~Koenker, \emph{Quantile Regression}, Cambridge University Press, Cambridge
  (2005),
  \href{https://doi.org/10.1017/CBO9780511754098}{10.1017/CBO9780511754098}.

\bibitem{kingma2017}
D.P.~Kingma and J.~Ba, \emph{Adam: a method for stochastic optimization},  in
  \emph{3rd International Conference on Learning Representations (ICLR 2015)},
  (San Diego, CA, USA), 2015 [\href{https://arxiv.org/abs/1412.6980}{{\ttfamily
  1412.6980}}].

\bibitem{bengio2009}
Y.~Bengio, J.~Louradour, R.~Collobert and J.~Weston, \emph{Curriculum
  learning},  in \emph{Proceedings of the 26th Annual International Conference
  on Machine Learning}, (Montreal, Canada), pp.~41--48, ACM, 2009,
  \href{https://doi.org/10.1145/1553374.1553380}{https://doi.org/10.1145/1553374.1553380}.

\end{thebibliography}\endgroup

\end{document}